\documentclass[a4paper]{jpconf}
\usepackage{graphicx}
\begin{document}
\title{A Progress Report on the Carbon Dominated Atmosphere White Dwarfs}

\author{P. Dufour$^1$, J. Liebert$^1$, B. Swift$^1$, G. Fontaine$^2$ and T. Sukhbold$^1$}

\address{$^1$ Steward Observatory, University of Arizona, 933 North Cherry Avenue, Tucson, AZ 85721}

\address{$^2$ D\'epartement de Physique, Universit\'e de Montr\'eal, C.P.6128, Succ. Centre-Ville, Montr\'eal, Qu\'ebec, Canada H3C 3J7}

\ead{dufourpa@as.arizona.edu}
 
\begin{abstract}
Recently, Dufour et al. (2007) reported the unexpected discovery that
a few white dwarfs found in the Sloan Digital Sky Survey had an
atmosphere dominated by carbon with little or no trace of hydrogen and
helium. Here we present a progress report on these new objects based
on new high signal-to-noise follow-up spectroscopic observations
obtained at the 6.5m MMT telescope on Mount Hopkins, Arizona.

\end{abstract}

\section{Introduction}

White dwarfs have traditionally been divided into two main categories
depending on the composition of the primary constituent of their
atmosphere. Stars with a hydrogen rich surface composition are
classified as DA white dwarfs while those with a helium dominated
surface composition belong to the non-DA category (DO, DB, DC, DZ and
DQ spectral type). A third category was recently uncovered by Dufour
et al. (2007) when an analysis of Hot DQ white dwarfs from the
Sloan Digital Sky Survey (SDSS) revealed that their surface composition
was dominated by carbon and not helium as was initially assumed
by Liebert et al. (2003).

These strange objects are quite rare. Only 9 were found in the SDSS
DR4 catalog of spectroscopically identified white dwarfs which
contained nearly 10,000 stars (Eisenstein et al. 2006). A detailed
analysis of the spectroscopic and photometric data by Dufour et
al. (2008) showed that the hot DQ white dwarfs are found in a rather
narrow range of effective temperature centered around $\sim$ 20,000 K.

No carbon dominated atmosphere white dwarfs with an effective
temperature higher than $\sim$ 24,000 K have been found from a careful
inspection of thousands of spectra in SDSS DR6 with $u-g < -0.2$ and
$g-r < -0.3$. The absence of hotter counterparts for carbon dominated
atmosphere white dwarfs suggest that they probably cool as helium
atmosphere stars until a convection zone develops in the underlying
carbon-rich mantle due to the recombination of that element. If this
scenario is correct, a subphotospheric carbon convection zone would
develop around 24,000 K and dilute the surface helium layer in the
much more massive carbon envelope. This would transform a helium
dominated DB star with a very thin surface helium layer into a carbon
dominated Hot DQ white dwarf (see Dufour et al. 2008 for more on this
scenario).

Why would some helium rich white dwarfs have thin helium layers is
still not entirely understood yet. One possibility is that these stars
are the result of the evolution of massive stars that have burned
carbon and have an oxygen-magnesium-neon core (see Garcia-Berro et
al. 1997 and reference therein). Alternatively, they could have
experienced a late thermal pulse that eliminated most of the helium, a
phenomenon similar to the one that is generally believed to explain
the existence of other hydrogen deficient stars (Werner \& Herwig
2006). There might also be other ways to eliminate the superficial
helium layers that still need to be explored. However, a scenario
involving close binary interaction, as proposed to explain the
variability of J1426+5752 (Montgomery et al. 2008), now appears very
unlikely since i) this star has been found to show multi-periodic
variations (Green et al. 2008, see also Fontaine et al. in these
proceedings) ii) no radial velocity variations have been detected in
J1426+5752 (see again Green et al. 2008) and iii) more importantly, it
doesn't explain why {\it all} Hot DQ's are found within the same
temperature range and {\it none} at higher or lower effective
temperatures. We thus believe that the convective mixing scenario
presented in Dufour et al. (2008) is the most appealing.

One essential parameter that would give important clues about the origin
and evolution of these objects is the mass. The mass of a white dwarfs is
generally obtained from a determination of the surface gravity by
matching the observed line profiles with those of model atmosphere
calculations. However, in the case of the Hot DQ white dwarfs, the
surface gravity is poorly known due to several factors.

First, the atmospheric parameter determinations of Dufour et al. (2008)
were based on the analysis of rather noisy optical SDSS spectra. Years
of experience from analyzing DA and DB white dwarfs has taught us that
high signal-to-noise ratio spectroscopic observations are required for
a precise measurement of the surface gravity (and effective
temperature). Rough estimates of log g were obtained by Dufour et
al. (2008) but the quality of the SDSS spectra is clearly insufficient
to provide the needed accuracy.

Second, most of the flux emitted by these stars is in the ultraviolet
part of the electromagnetic spectrum. This region of the spectrum
contains many more absorption features than there is in the
optical. All these absorption lines have an important influence on the
thermodynamic structure (temperature and pressure as a function of
optical depth) of the atmosphere since the flux absorbed at short
wavelength tends to be redistributed at longer wavelength (see Fig. 4
of Dufour et al. 2008). Hence, small errors in our modeling of the
UV part of the spectrum will have an important effect on the line
profiles in the optical. Unfortunately, an assessment of the quality
of our modeling of the UV will only be possible by comparing UV
observations (which we should obtain, if the HST repair mission goes as
planed, sometime next year) with our model calculations.

Finally, any determination of the surface gravity from the line
profiles must rely on a state-of-the-art theory of the broadening of
the spectral lines. At the time of the Dufour et al. (2008) analysis,
detailed calculations for the Stark broadening of CII lines were not
available and the standard scaled classical approximation was
used. Therefore, surface gravities (and effective temperatures)
obtained in Dufour et al. (2008) can only be considered as preliminary
approximations until new tables for the Stark broadening of CII (which
are currently being calculated) are incorporated in our next
generation of model atmosphere.

In summary, atmospheric parameters will only be truly reliable when each of
the three issues above have been properly addressed. This paper
report our development on the first of these issues.

\section {New MMT Observations} 

\begin{figure}
\begin{center}
\includegraphics[width=6in]{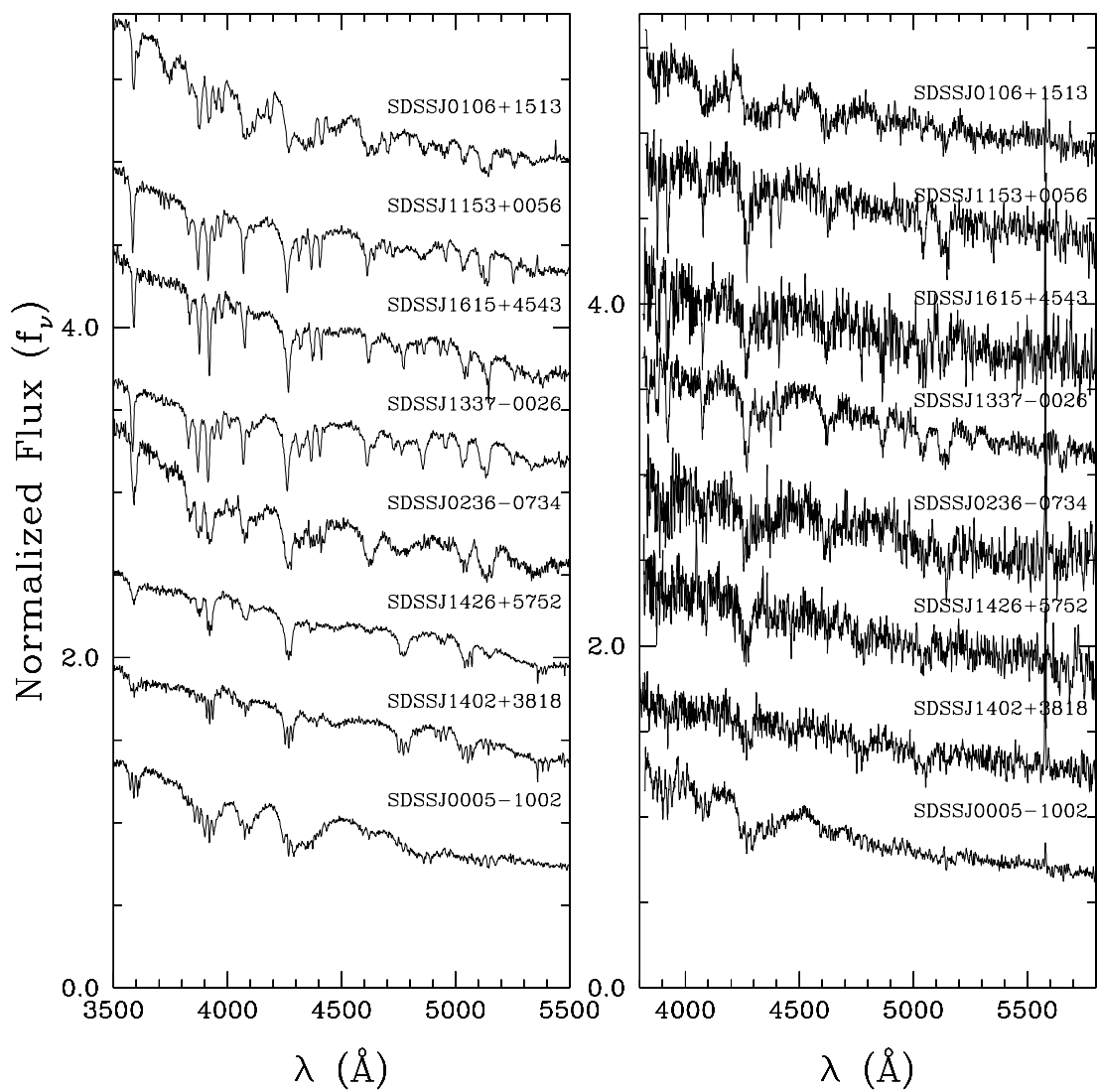}
\end{center}
\caption{Left: New spectroscopic observations from the 6.5m MMT telescope for 8 of the 10 known 
Hot DQ white dwarfs. Spectra are normalized to unity at 4500 \AA~ and
offset from each other for clarity. The stars are approximately
ordered by increasing effective temperature from bottom to top. Right:
Corresponding SDSS spectra for each of the Hot DQ that have been
reobserved with MMT. Note that we have applied for clarity a
three-point average window smoothing in the display of the SDSS
spectroscopic data.}
\end{figure}

As mentioned above, in order to obtain a precise determination of the
atmospheric parameters, it is of utmost importance to have high
signal-to-noise ratio spectroscopic observations. Due to the nature of
the Sloan Digital Sky Survey acquisition procedure, spectra of faint
objects such as the Hot DQ are of rather poor quality. We have thus
recently undertaken a program to reobserve all the known carbon
dominated atmosphere white dwarfs with the Mt. Hopkins 6.5m MMT
telescope. We have obtained 7 nights of MMT time and succeeded in
securing high signal-to-noise ratio spectra (typically S/N of 60 or
more) for 8 of the 10 (9 from the DR4 WD catalog and a 10th one
recently found in DR6) known Hot DQ white dwarfs (see Figure 1). We
used the Blue Chanel with the 500 line mm$^{-1}$ grating with a 1''
slit, resulting in a $\sim$ 3.6 \AA~FWHM spectral resolution at 5500
\AA. The spectra were reduced with standard IRAF packages. The setup
used allowed a coverage of a few hundred angstroms bluer than that of
SDSS although at the expense of an equally good coverage of the redder
part of the electromagnetic spectrum (our spectra cover a wavelength
range of $\sim$3400-6300 \AA~ vs $\sim$3800-9000 \AA~ for SDSS).

In Figure 1, we present the new high signal-to-noise ratio spectrum
for all the Hot DQ that have been reobserved with MMT so far. We also
show, for comparison, the corresponding SDSS spectra that were used
for the preliminary analysis of Dufour et al. (2008). The noise level
and sharpness of the observed spectral lines are now suitable for an
eventual precise spectroscopic analysis. Howerver, it would be meaningless at
this point to present updated atmospheric parameters based on these
new observations since our grid does not incorporate the new tables
for the Stark broadening of CII yet. We prefer to wait for the next
generation of models, and also the HST observations, before sharing
the results of our reanalysis for these stars. Nevertheless, even
without proceeding with a full detailed analysis, several interesting
observations are worth discussing.

The most remarkable concerns the fraction of Hot DQ stars that are
found to be magnetic. In Figure 1, we clearly see magnetic splitting of
the spectral lines for at least 4 objects. These are the 4 objects
starting from the the bottom in Figure 1. The separation between the
$\sigma$ and $\pi$ components indicates that the fields are of the order of
1 MG. The splitting was not observable in the noisier SDSS spectra
except for the brightest Hot DQ SDSS J0005-1002. Spectropolarimetric
measurement has also confirmed the magnetic nature for SDSS J0005-1002
(see Dufour et al. 2008). This is in contrast with the next 3 stars in
Figure 1 that show instead very sharp CII lines where no trace of
magnetic splitting is detectable. Given the spectral resolution of our
observations, a limit of about 200 kG can be put on the strength of
the magnetic field for these objects. The hottest stars in our sample,
SDSS J0106+1513, do not show sharp and well defined lines which
suggest perhaps the presence of a weak magnetic field but more precise
spectropolarimetric measurements will be needed to confirm
magnetism. Note also that one of the two stars that remains to be
observed with MMT, SDSS J2200-0741, seems also to have lines slightly
broadened by a weak magnetic field.

It thus seems that at least 4 (and perhaps 6) of the 10 known Hot DQ
are magnetic. This is an incredibly high fraction considering that the
fraction of magnetism of nearby white dwarfs is around 10$\%$ (Liebert
et al. 2003). Is this an indication that Hot DQ have
higher mass than average since magnetic white dwarfs tend to be more
massive in general? The preliminary analysis of Dufour et al. (2008)
indicated that most of these stars have surface gravities around log g
$\sim$ 8.0 although this could possibly be revised if it is found
that our approximation for the Stark broadening considerably
overestimated the broadening of the lines.

This unusual high fraction could indicate that magnetism plays a
significant role in the evolution of these objects. But if this is the
case, why do we detect magnetism in only about 50$\%$ and not 100$\%$
of these stars? Could it be that they are indeed all magnetic but that
in some cases the strength of the field is not strong enough to be
detected from line splitting? More sensitive spectropolarimetric
measurements, which we plan to obtain soon, should elucidate those
questions.

We also detect the HeI 4471 line, indicating the presence of a
significant amount of helium (C/He $\sim$0.5), in two of the coolest
Hot DQ (SDSS J1426+5752 and SDSS J1402+3818). The presence of helium
in SDSS J1426+5752 was also needed to explain the pulsational
properties recently discovered by Montgomery et al. (2008) since pure
carbon atmosphere white dwarfs are not expected to pulsate at this
temperature (see Fontaine et al. 2008 and Fontaine et al., these
proceedings).

\begin{figure}
\begin{center}
\includegraphics[width=6in]{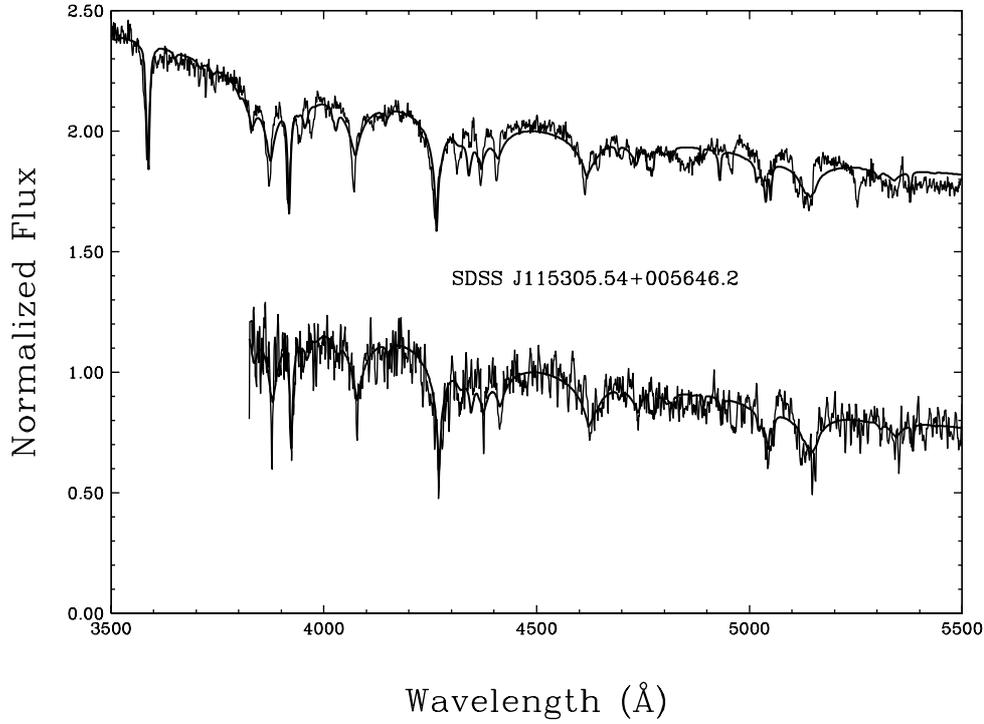}
\caption{Comparison of the fits of the SDSS (bottom) and MMT (top) spectra for SDSS J1153+0056. The solution is the same as that presented in Dufour et al. 2008, that is log =8 and $T_{\rm eff}$ = 21,650 K.}
\end{center}
\end{figure}

We will conclude this discussion with some remarks on our models. The
solutions presented in Dufour et al. 2008 were not bad considering
the level of noise of SDSS spectra.  However, with the new MMT
observations, it is now obvious that improvement in our models will be
needed before we can claim that we know the atmospheric parameters
with great accuracy. This is illustrated in Figure 2 where we show the
Dufour et al. 2008 solution and a very similar fit to the MMT data for
SDSS J1153+0056. While the fit to the SDSS data looks quite
acceptable, there are significant discrepancies that appear when
observing in detail the MMT fit. It seems that our models do not
reproduce correctly the strength of many lines. This is most evident
when looking at the group of lines near 4400-4500 \AA (and it can also
be observed at many other wavelength). Some of the lines are quite
well reproduced with our models while others seem to require a
different log g or $T_{\rm eff}$ (see Figure 1 of Dufour et al. 2008
for example). This is usually symptomatic of a wrong thermodynamic
structure. This could be due, as already mentioned above, to an
improper modeling of the UV part of the spectrum. The treatment of
the Stark broadening of CII lines might also cause such a
discrepancy. The exact reason is probably a combination of both, as
well as other parameters such as the abundance of other trace
elements, treatment of the convective efficiency or perhaps even
uncertainties in the atomic data used. Carbon dominated atmosphere
modeling is relatively new and has obviously not reached yet the level
of accuracy that is attained for DA and DB stars. We hope to achieve
this in a near future.

\section {Conclusions} 

White dwarfs with a carbon dominated atmosphere represent a new
challenge to stellar evolution. Their origin and evolution are very
mysterious and as such, they deserve all our attention. Progress on
our understanding of these objects will necessitate i) new models with
state-of-the-art physics (Stark broadening) ii) observation in the UV
part of the electromagnetic spectrum (HST observations are coming
soon, assuming no problems with the next repair mission), and iii)
high signal-to-noise ratio spectroscopic data. In this short report,
we presented new optical spectra taken with MMT. Their analysis will
eventually lead to a more precise measurements of the atmospheric
parameters. One of the most outstanding characteristics that emerges from
these new observations is the unusually high fraction of Hot DQ with a
large magnetic field. Weather or not the magnetic field is intimately
related to their evolution and origin remains unclear. Undoubtedly, 
further study of Hot DQ white dwarfs will increase our knowledge of
stellar evolution. The new challenges that they bring to our field will
be of the most interesting to solve.

\subsection*{Acknowledgments}
P.D. acknowledges the financial support of NSERC. This work was also
supported by the National Science Foundation through grant AST
03-07321. Support from the Reardon Foundation is also gratefully
acknowledged.

\section*{References}

\begin{thereferences}

\item Dufour, P., Fontaine, G., Liebert, J., Schmidt, G.~D., \& Behara, N.\ 2008, {\sl ApJ}, {\bf 683}, 978 

\item Dufour, P., Liebert, J., Fontaine, G., \& Behara, N.\ 2007, {\sl Nature}, {\bf 450}, 522 

\item Eisenstein, D. J. et al.\ 2006, {\sl ApJ}, {\bf 167}, 40

\item Fontaine, G., Brassard, P., \& Dufour, P.\ 2008, {\sl A\&A}, {\bf 483}, L1 

\item Garcia-Berro, E., Ritossa, C., Iben, I. Jr. \ 1997 {\sl ApJ}, {\bf 485}, 765

\item Green, E. M., Dufour, P., Fontaine, G. in preparation

\item Liebert, J. et . al.\ 2003, {\sl AJ}, {\bf 126}, 2521

\item Liebert, J., Bergeron, P., Holberg, J. B. \ 2003 {\sl AJ}, {\bf 125}, 348

\item Montgomery, M.~H., Williams, K.~A., Winget, D.~E., Dufour, P., DeGennaro, S., \& Liebert, J.\ 2008, {\sl ApJL}, {\bf 678}, L51 

\item Werner, K., Herwig, F. \ 2006 {\sl PASP}, {\bf 118}, 183

\end{thereferences}

\end{document}